\documentclass{aa}
\newcommand{\EQ}{\begin{equation}}
\newcommand{\EN}{\end{equation}}
\newcommand{\EQA}{\begin{eqnarray}}
\newcommand{\ENA}{\end{eqnarray}}
\newcommand{\eq}[1]{(\ref{#1})}

\newcommand{\Eq}[1]{Eq.~(\ref{#1})}
\newcommand{\Eqs}[2]{Eqs.~(\ref{#1}) and~(\ref{#2})}

\newcommand{\Eqss}[2]{Eqs.~(\ref{#1})--(\ref{#2})}

\newcommand{\Sec}[1]{Sect.~\ref{#1}}

\newcommand{\Fig}[1]{Fig.~\ref{#1}}

\newcommand{\Figs}[2]{Figs~\ref{#1} and \ref{#2}}

{}
{}
{}
{}
{}
{}
{}
\newcommand{\uu}{{\vec{u}}}

\newcommand{\ff}{\mbox{\boldmath $f$} {}}

\newcommand{\FF}{{\vec{F}}}

\newcommand{\nab}{\vec{\nabla}}

\newcommand{\SSSS}{\mbox{\boldmath ${\sf S}$} {}}

\newcommand{\DDD}{{\cal D} {}}

\def\half{{\textstyle{1\over2}}}

\newcommand{\yaj}[3]{ #1, {AJ,} {#2}, #3}

\newcommand{\yapj}[3]{ #1, {ApJ,} {#2}, #3}
\newcommand{\yapjl}[3]{ #1, {ApJ,} {#2}, #3}

\newcommand{\yana}[3]{ #1, {A\&A,} {#2}, #3}

\newcommand{\ypf}[3]{ #1, {Phys. Fluids,} {#2}, #3}

\newcommand{\ymn}[3]{ #1, {MNRAS,} {#2}, #3}

\newcommand{\yicarus}[3]{ #1, {Icarus,} {#2}, #3}

\newcommand{\yjour}[4]{ #1, {#2}, {#3}, #4}

\usepackage{graphicx}
\usepackage{url}
\usepackage{amssymb}
\usepackage{epsfig}
\usepackage[english]{babel}

\DeclareMathSymbol{\varOmega}{\mathord}{letters}{"0A}
\DeclareMathSymbol{\varPsi}{\mathord}{letters}{"09}
\DeclareMathSymbol{\varPhi}{\mathord}{letters}{"08}

\begin{document}
\authorrunning{A.\ Johansen et al.}
\titlerunning{Dust-trapping vortices in protoplanetary discs}
\title{Simulations of dust-trapping vortices in protoplanetary discs}
\author{Anders Johansen\inst{1,}\inst{2}
        \and
        Anja C.\,Andersen\inst{1}
        \and
        Axel Brandenburg\inst{1}}
\institute{NORDITA, Blegdamsvej 17, DK-2100 Copenhagen, Denmark \\
        \email{ajohan/anja/brandenb@nordita.dk}
         \and
         Astronomical Obs., NBIfAFG, Copenhagen University,
     Juliane Maries Vej 30, DK-2100 Copenhagen, Denmark\\
}\date{\today,~ $ $Revision: 1.2 $ $}

\abstract{
Local three-dimensional shearing box simulations of the compressible
coupled dust-gas equations are used in the fluid approximation to study
the evolution of different initial vortex configurations in a protoplanetary
disc and their
dust-trapping capabilities.
The initial conditions for the gas are derived from an analytic solution
to the compressible Euler equation and the continuity equation. The solution is valid if there is
a vacuum outside the vortex.
In the simulations the vortex is either embedded in a hot corona,
or it is extended in a cylindrical fashion in the vertical direction.
Both configurations are found to survive for at least one orbit and
lead to accumulation of dust inside the vortex.
This confirms earlier findings that dust accumulates in anticyclonic vortices, indicating that this is a viable
mechanism for planetesimal formation.

}

\maketitle

\section{Introduction}

The expected scenario for planet formation within a protoplanetary disc around
a newly formed star is that planets grow from kilometre-sized planetesimals to
protoplanets by gravitationally attracting each other. The planetesimals are
expected to form within the protoplanetary disc while gas is still present
in the disc. The coupling of gas and solids is therefore an important issue in
the formation of planets.

Planetesimals can only grow from sub-micron solids if their
relative velocities are less than about 1~ms$^{-1}$, since larger
velocities will result in disruption of the aggregate (Blum \& Wurm
\cite{blum+wurm00}). Relative velocities considerably higher than 1~ms$^{-1}$
arise from chaotic motions in a turbulent disc or from differential orbital
drift in a laminar disc (Weidenschilling \& Cuzzi
\cite{weidenschilling+cuzzi93}). Another problem facing planetesimal growth is
the rapid inward orbital drift associated with gas drag that carries small
bodies into the star on a time scale of 100-1000 years
(Weidenschilling \cite{weidenschilling77}).

One attractive scenario which may be able to overcome the two above problems
is the presence of dust-trapping anticyclonic vortices within the
protoplanetary disc (Barge \& Sommeria \cite{barge+sommeria95}; Tanga
et al.\ \cite{tanga+etal96}; Bracco et al.\ \cite{bracco+etal99}; Godon \&
Livio \cite{godon+livio00}). If dust is trapped within vortices, the
relative velocities between solid particles would be small since
all the solids rotate in the same direction within the vortex. Trapping would
also prevent solid bodies from spiralling inwards towards the
star. Self-gravity between sufficiently large amounts of trapped
solid material inside vortices could
even trigger a local gravitational instability and subsequent growth of
centimetre-sized solid bodies into planetesimals.

The formation and stability of vortices has been an area of much 
research since the dust-trapping mechanism was proposed by Barge \& Sommeria
(\cite{barge+sommeria95}). Using a two-dimensional incompressible model, Bracco et al.\
(\cite{bracco+etal98}; \cite{bracco+etal99}) show that anticyclonic vortices
can form as a relic from the originally strongly turbulent disc.
The existence of a baroclinic instability in a protoplanetary disc is investigated by Klahr \&
Bodenheimer (\cite{klahr+bodenheimer03}). This instability works as a source of
vorticity and forms vortices similar to how vortices form around high and
low pressures on Earth. The stability of vortices is simulated by Godon \&
Livio (\cite{godon+livio99}). They show that two-dimensional anticyclonic
vortices can survive for hundreds of orbits, and that isolated vortices can
merge to form larger vortices.

The dust-trapping mechanism was investigated by Hodgson \& Brandenburg
(\cite{hodgson+brandenb98}) in three dimensional simulations of disc turbulence
driven by the magneto-rotational instability (Balbus \& Hawley
\cite{balbus+hawley1998}). In these simulations the vortices are highly
time dependent, but when gas velocity is frozen in time, they find
concentration of dust inside vortices. However, when considering a freely
evolving gas flow, they no longer see any significant dust concentration
inside vortices. This is associated both with the finite life time of
turbulent vortices and with the stirring up of dust by turbulence in the
vertical direction.
The dust-trapping efficiency of vortices was explored analytically by
Chavanis (\cite{chavanis00}) for vortices of arbitrary aspect ratio.
Recently, de la Fuente Marcos \& Barge
(\cite{fuente+barge01}) have considered a single frozen vortex velocity field in two
dimensions and a distribution of particle sizes. Using realistic expressions
for the drag force on dust they find efficient particle trapping
inside vortices, preventing the inwards orbital drift associated with gas drag.

In the present paper we explore the dust-trapping efficiency of three-dimensional
vortices. We start out from the ana\-lytic vortex solution by Goodman et al.\
(\cite{goodman+etal87}). This solution takes density and
pressure gradients across the vortex explicitly into account. We let the gas
flow evolve freely and exa\-mine the coherence of the three-dimensional
vortices and their ability to trap dust.

The paper is structured as follows. In \Sec{s_bashydeq} we give the basic
hydrodynamic equations used. In \Sec{s_initialcond} we discuss in detail the
vortex solution that we use as initial condition. The dust-trapping mechanism is described in
\Sec{s_dusttrapping}. In \Sec{s_3dmodels} we discuss the two models that
we use. In \Sec{s_numericalmethod} we describe the numerical scheme and the
boundary conditions implemented. Our results are presented in \Sec{s_results}.

\section{Basic hydrodynamic equations}
\label{s_bashydeq}

We perform simulations in the shearing sheet approximation (Wisdom \& Tremaine
\cite{WisdomTremaine88}, Hawley et al.\ \cite{hawley+gammie+balbus95}), where
a local coordinate frame corotating with the Kepler flow at a
distance $r_0$ from the central star is considered. In this local approximation the
curvilinearity of the coordinates is neglected, so the validity is limited to
the case when the size of the vortex is much smaller than $r_0$. The $x$-axis
points away from the star, and the $y$-axis points along the flow direction. The
angular velocity profile of a Keplerian disc goes as $\varOmega(r)\propto r^{-q}$,
where $q=1.5$ when considering only the gravitational attraction to the central
star. 

In the coordinate frame both inertial and fictitious forces exist. Radial
gravity and centrifugal force terms only cancel at the radius $r_0$, giving
rise to a tidal force away from $r_0$.  However, when measuring velocities
relative to the main shear flow, $\uu^{(0)}\equiv(0,-q\varOmega x,0)$, this tidal
force vanishes. We then have $\vec{u} = \vec{u}^{(0)} + \tilde{\vec{u}}$ and use
$\tilde{\vec{u}}$ as the velocity variable. The vertical part of the gravity still gives rise to a
restoring force proportional to $-\varOmega^2 z$ in the $z$-direction. Gas
particles also experience a pressure gradient force, viscous forces and the
fictitious Coriolis force. 
 
We treat gas and dust as two separate fluids and let the two interact
through a drag force. We use the term dust for solid bodies of all sizes. The drag force from gas upon the dust attempts to
accelerate dust to match the velocity of the gas, and vice versa. We
assume that dust is not pressure supported (due to a low number density
and low particle velocities).  
Since the solid density of dust particles is many orders of magnitude higher
than the gas density and the pressure gradient acts as a volume force, it is
reasonable to assume that dust is not affected by gas pressure (see e.g.\
Seinfeld \cite{seinfeld86}).

\subsection{The shearing sheet approximation}
\label{s_shearing}

In the shearing sheet approximation the equations for the departure from the
main shear flow (Brandenburg et al.\ \cite{BNST95}) can be written in the form
\EQ
  \rho\left[{\DDD\tilde{\vec{u}}\over\DDD t}
  +\tilde{\vec{u}}\cdot\nab\tilde{\vec{u}}-\ff(\tilde{\vec{u}}) \right]  \!=\!
  -\nab P + \vec{F}_{\rm visc} -\beta(\tilde{\vec{u}}-\tilde{\vec{u}}_{\rm d})
  \, ,
\EN
\EQ
  \rho_{\rm d}\left[{\DDD\tilde{\vec{u}}_{\rm d}\over\DDD t}
  +\tilde{\vec{u}}_{\rm d}\cdot\nab\tilde{\vec{u}}_{\rm
  d}-\ff(\tilde{\vec{u}}_{\rm d})  \right] \!=\! \vec{F}_{\rm visc, d} -
  \beta(\tilde{\vec{u}}_{\rm d}-\tilde{\vec{u}}),
\EN
\EQ
  {\DDD\rho\over\DDD t}+\nab\cdot(\rho\tilde{\vec{u}})=0,
  \label{contgas}
\EN
\EQ
  {\DDD\rho_{\rm d}\over\DDD t}+\nab\cdot(\rho_{\rm d}\tilde{\vec{u}}_{\rm d})=0,
\EN
\EQ
  \rho T\left({\DDD s\over\DDD t}+\tilde{\vec{u}}\cdot\nab
  s\right)=2\rho\nu\SSSS^2 + \rho \zeta (\nab \cdot \tilde{\vec{u}})^2
  + \nab \cdot (K \nab T) \, ,
  \label{dsdt}
\EN
where $\tilde{\vec{u}}$, $P$, $\rho$, $T$ and $s$ are the velocity, pressure,
density, temperature and specific entropy of gas,
respectively, and $\tilde{\vec{u}}_{\rm d}$ and $\rho_{\rm d}$ are the velocity and density
of dust. The density of dust is defined as $\rho_{\rm d} = n_{\rm d}
m_{\rm d}$, where $n_{\rm d}$ is the number density of dust particles and
$m_{\rm d}$ is their mass (all particles are assumed to have the same mass).
The parameter $\zeta$ describes some bulk viscosity, see \Sec{s_viscosity}.
Dust density is not to be confused with the solid density $\rho_{\rm s}$
of individual dust particles (defined in \Sec{s_stoppingtime}). The advective
derivative, $\DDD/\DDD t=\partial/\partial t+u_y^{(0)}\partial/\partial y$, is
with respect to the mean shear flow only,
\EQ
  \ff(\tilde{\vec{u}})=\pmatrix{2\varOmega \tilde{u}_y\cr-\half\varOmega
  \tilde{u}_x\cr-\varOmega^2z}
\EN
is the Coriolis force combined with the tidal force and the vertical
gravity force, $\vec{F}_{\rm visc}$ and $\vec{F}_{\rm visc, d}$ are viscosity
forces (see next section) and $\beta$ is the coupling coefficient
between gas and dust. This coupling coefficient can be expressed in terms of
the stopping time $\tau_{\rm s}$, which in turn is defined through the relation
\EQ
  \FF_{\rm D}=-\frac{\rho_{\rm d}}{\tau_{\rm s}} (\tilde{\vec{u}}_{\rm
  d}-\tilde{\vec{u}})    
  \, ,
  \label{draglaw}
\EN
where $\FF_{\rm D}$ is the drag force per unit volume, so $\beta = \rho_{\rm
d}/\tau_{\rm s}$. The stopping time is a parameter that describes the
interaction between a single dust particle and the surrounding gas and does not depend on the number density of
dust particles, whereas $\beta$ does. 

Temperature, pressure and density are related to each other by the
relation $P=(c_{\rm p}-c_{\rm v})T\rho$, where $c_{\rm p}$ and $c_{\rm v}$ are the specific
heats at constant pressure and constant volume. The adiabatic sound speed
$c_{\rm s}$ is given by
\EQ
  c_{\rm s}^2=c_{\rm s0}^2\exp\left[\gamma s/c_{\rm p} +
  (\gamma-1)\ln(\rho/\rho_0)\right],
\EN
where $\rho_0$ and $c_{\rm s0}$ are arbitrarily chosen integration constants
from the integration of the first law of thermo\-dynamics
and $\gamma \equiv
c_{\rm p}/c_{\rm v}$ is the ratio of specific heats at constant pressure and volume,
respectively. We have chosen integration constants such that $s=0$ when 
$c_{\rm s}=c_{\rm s0}$ and $\rho=\rho_0$.
The thermal conductivity $K$ is related to the kinematic (shear) viscosity $\nu$
through the non-dimensional Prandtl number, $\mbox{Pr}=\nu/(\rho c_{\rm p} K)$.

\subsection{Viscosity}
\label{s_viscosity}

The viscosity force on the gas can be expressed as
\EQ
  {\vec{F}_{\rm visc}\over\rho}\! =\! \nu \left( \nab^2 \tilde{\vec{u}} 
      + {\textstyle\frac{1}{3}}\nab \nab \cdot \tilde{\vec{u}} 
      + 2 \vec{\mathsf{S}} \cdot \nab \ln \rho \right)
      + \zeta \nab \nab \cdot \tilde{\vec{u}} ,
\EN
where $\nu$ is the kinematic viscosity of the gas (assumed constant),
\EQ
  \mathsf{S}_{ij} = \frac{1}{2} \left( \frac{\partial \tilde{u}_i}{\partial x_j} +
  \frac{\partial \tilde{u}_j}{\partial x_i} -\frac{2}{3} \delta_{ij} \nab \cdot
  \tilde{\vec{u}}\right)
\EN
is the traceless rate-of-strain tensor and $\zeta$ is the bulk
viscosity, which is invoked solely to smear out sharp gradients (or shocks)
over a few mesh points near strongly con\-verging regions.
The shock viscosity, $\zeta=\zeta_{\rm shock}$, as it is used here,
is proportional to the smoothed maximum
of the positive flow convergence,
\EQ
  \zeta_{\rm shock}=c_{\rm shock}\left<\max_5[(-\nab \cdot \tilde{\vec{u}})_+]\right>,
\EN
where $c_{\rm shock}$ is a non-dimensional coefficient
defining the strength of the shock viscosity.
The smoothing is accurate to second order, and the maximum is taken 
over the second-nearest points.

The viscous force on the dust would normally be negligibly small, but when
treating dust as a fluid one always has to add a small viscosity to
damp out unphysical wiggles on the mesh scale. It proved sufficient to use
\EQ
  \rho_{\rm d}^{-1}\vec{F}_{\rm visc, d}=\nu_{\rm d} \nab^2
  \tilde{\vec{u}}_{\rm d},
\EN
where we have ignored any dependence on dust density, because
$\vec{F}_{\rm visc, d}$ is small anyway.

\subsection{The stopping time}
\label{s_stoppingtime}

Due to Newton's third law the value of $\beta$ must be the same for dust and
gas, which leads to
\EQ
  \frac{\rho_{\rm d}}{\tau_{\rm s}} = \frac{\rho}{\tau_{\rm s,g}} \, .
  \label{tstopgd}
\EN
Specifying the stopping time of dust automatically sets the stopping time of
gas, $\tau_{\rm s,g}$, according to \Eq{tstopgd}. The stopping time of gas
decreases with increasing dust density. This means that the back-reaction of
dust upon gas should become more and more important as dust density
approaches gas density inside a vortex.

A non-dimensional measure of the stopping time in terms of the angular
velocity is the parameter $\varOmega \tau_{\rm s}$.
If $\varOmega \tau_{\rm s}$ is small, then dust is strongly
coupled to gas, whereas when $\varOmega \tau_{\rm s}$ is large, dust is
relatively unaffected by gas drag.

Analytic expressions for the stopping time of dust due to gas drag exist in
various regimes, depending on the mean free path of the particle and the
Reynolds number of the flow (e.g.\ Weidenschilling
\cite{weidenschilling77}; see also
Chavanis \cite{chavanis00}; de la Fuente Marcos \& Barge
\cite{fuente+barge01}). The dust particles are assumed spherical with radius
$a_{\rm s}$, solid density $\rho_{\rm s}$ and mean free path $\lambda$. If $\lambda <
\frac{4}{9} a_{\rm s}$, the Stokes drag law is valid. Here the stopping time
depends on Reynolds number $\mbox{Re}$. When $\lambda > \frac{4}{9} a_{\rm s}$, one
must use the Epstein drag law. The linear drag law used here, where
$\tau_{s}$ is constant in \Eq{draglaw}, is only valid when the stopping time is
independent of relative velocity. This is the case in two regions: in the
Stokes drag regime with $\mbox{Re}<1$ and in the Epstein drag regime (see
Weidenschilling \cite{weidenschilling77}). The transition between Epstein and
Stokes regime in the Solar nebula depends on dust particle radius (Chavanis
\cite{chavanis00}). In the outer Solar System particles will typically be in
the Epstein regime, so the drag law used here is valid for the outer Solar
System, i.e.\ from the location of Jupiter and outwards. We ignore the
dependence of $\tau_{\rm s}$ on local gas density.

In the Epstein drag regime there is a simple expression for the stopping time, 
\EQ
  \tau_{\rm s} = \frac{\rho_{\rm s}}{\rho} \frac{a_{\rm s}}{c_{\rm s}} \, .
\EN
To calculate the radius of a particle whose stopping time is known, we consider
a typical protoplanetary disc with a scale height $H=10^{12} \, \rm{cm}$
and a mass ratio $\rho_{\rm s}/\rho = 10^{10}$ in the mid-plane. The scale
height of the disc can be expressed in terms of the speed of sound and the Kepler
frequency under the assumption of vertical hydrostatic equilibrium and an
isothermal density profile,
\EQ
  H = {c_{\rm s}}/{\varOmega} \, ,
\EN
so
\EQ
  a_{\rm s} = \frac{\rho}{\rho_{\rm s}} H \varOmega \tau_{\rm s} = 10^2 \,    
  \varOmega \tau_{\rm s} \, \rm{cm} \, .
  \label{taustorad}
\EN

\subsection{Settling of dust}
\label{s_dustsettling}

Dust is subjected to a gravity force in the vertical direction without a
balancing pressure gradient force, which makes it fall towards the mid-plane.
The terminal velocity (the velocity where gravity and drag force
balance) at height $z$ is determined by the stopping time as
\EQ
  0 = -\varOmega^2 z - \frac{1}{\tau_{\rm s}} u_{{\rm d}z}^{\rm (ter)} \, ,
\EN
so $u_{{\rm d}z}^{\rm (ter)} = -\tau_{\rm s} \varOmega^2 z$.
At $z=H$ the terminal speed reaches the fraction $\tau_{\rm s} \varOmega$
of the sound speed, in which case the linear drag law \Eq{draglaw} breaks down
for bodies with large stopping times.

For long stopping times, $\tau_{\rm s} > \varOmega^{-1}$, dust will essentially
free-fall towards the mid-plane. The solid particles then perform damped
oscillations around the mid-plane and eventually settle to form a thin
sheet around $z=0$.
This was long considered the seed of planetesimals: the thin sheet will become
gravitationally unstable and then planetesimals will condense out (Safronov
\cite{safronov69}; Goldreich \& Ward \cite{goldreich+ward73}). However,
today it is believed that turbulence caused by the shear between
the thin dust layer and the gas leads to a continuous stirring up
of the dust (Weidenschilling \cite{weidenschilling80}, Cuzzi et
al.\ \cite{cuzzi+etal93}). In our present model the first arrival of the solid
particles at the mid-plane (forming essentially a delta-function) leads to the
dust density being under-resolved for given resolution.
We avoid this problem by ignoring the vertical gravity on the dust
thereby maintaining the initial scale height.

\section{Initial conditions}
\label{s_initialcond}

\subsection{The vortex solution}
\label{ss_vortexsolution}

GNG showed that there exists
an elliptic vortex flow solution to the Euler and continuity equations in the
shearing sheet approximation. In this
section we briefly describe their solution.
In Appendix A we go into more detail about how to arrive at this
solution.

\begin{figure}[]
  \begin{center}
    \epsfig{file=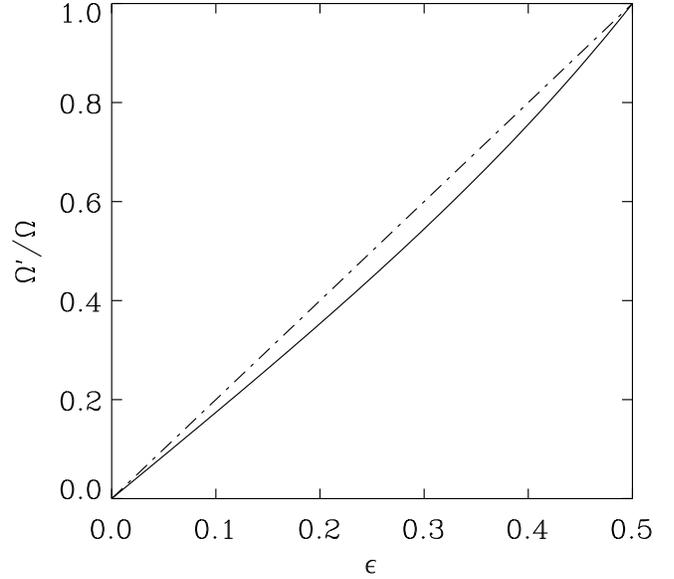}
  \end{center}
  \caption[]{The angular velocity of a vortex $\varOmega'$ as a function of
  aspect ratio $\epsilon$.
  The dash-dotted line connecting the end-points is present to illustrate the
  curvature.}
  \label{f_angvel}
\end{figure}

For a better distinction of the different forces involved we look at the
equations with the main shear flow included again. The Euler and continuity
equations for the motion of gas then take the explicit form (when ignoring
drag force and viscosity)
\EQA
  \frac{\partial \vec{u}}{\partial t} + (\vec{u} \cdot \nab) \vec{u} &=&
  \varOmega^2 (3 \vec{x} -\vec{z}) - 2 \vec{\varOmega}
  \times \vec{u} - \frac{1}{\rho} \nab P \, , 
  \label{eulgood} \\
  \frac{\partial \rho}{\partial t} &=& - \nab \cdot (\rho \vec{u}) 
  \, , \label{contgood}
\ENA
where $\vec{x}=(x,0,0)$ and $\vec{z}=(0,0,z)$. The $3\varOmega^2\vec{x}$ term
is the tidal force approximated along the $x$-direction to first order, and
the $-\varOmega^2\vec{z}$ term
is the vertical gravity force. The last two
terms on the right hand side of \Eq{eulgood} are the Coriolis force and the pressure gradient
force. We stress that
the equations in this form are completely similar to the Euler and continuity
equations of \Sec{s_shearing}. We introduce the specific enthalpy $h$ and write
$-\rho^{-1} \nab P = - \nab h$ in
the isentropic case. Then the GNG solution for the elliptical velocity field
and the corresponding enthalpy is
\EQA
  u_x &=&  \epsilon \varOmega' y \label{uxgood}\, , \\
  u_y &=& -\frac{1}{\epsilon} \varOmega' x \label{uygood}\, , \\
  u_z &=& 0 \, , \label{uzgood} \\
  h &=& \frac{1}{2} \delta^2 \varOmega^2 (b^2-x^2-\epsilon^2 y^2) - \frac{1}{2}
  \varOmega^2 z^2 \, , \label{hgood}
\ENA
where $\epsilon=a/b$ is the aspect ratio of the ellipse and $\varOmega'$ is the
angular velocity of the vortex. The ellipse has a semi-major axis $a$ in the
$y$-direction (along the Kepler flow) and a semi-minor axis $b$ in the
$x$-direction (radially outward). The aspect ratio must be in the interval $0
\leqslant \epsilon \leqslant 0.5$. The velocity field is incompressible and has
no $z$ component. The parameters $\varOmega'$ and $\delta$ are connected to
the aspect ratio of the ellipse and the background Keplerian angular velocity
through
\EQ
  \varOmega' = \alpha \varOmega = \frac{\sqrt{3} \epsilon}{\sqrt{1-\epsilon^2}}
  \varOmega
\EN
(see \Fig{f_angvel}), and
\EQ
  \delta^2 = -\frac{3}{1-\epsilon^2}+\frac{2 \sqrt{3}}{\sqrt{1 - \epsilon^2}}
  \, .
\EN
These two bindings are necessary for fulfilling the continuity equation (see
Appendix A). The solution is valid where $h \geqslant 0$, giving the vortex an
ellipsoidal shape with axes $a=b/\epsilon$, $b$ and $c=\delta b$.

The anticyclonic vortex feels a compressive Coriolis force, which is balanced
by pressure and tidal forces. The anisotropy of the latter gives rise to the
elliptical shape. In the pressureless limit, the streamlines reduce to Keplerian
epicycles. We explain the vortex dynamics and the dust-trapping mechanism in
\Sec{s_dusttrapping}.

\subsection{Units}
\label{s_units}

The model is scale-invariant. This means that the basic units can be chosen
arbitrarily. We take
\EQ
  [t]=\varOmega^{-1},\quad [x]=b, \quad [\rho]=\rho_0, \quad [s]=c_{\rm p}
  \, ,
\EN
where $b$ is the semi-minor axis of the vortex in the horizontal plane,
i.e.\ in the cross-stream direction, and $\rho_0$ is the average
density of gas in the whole box.
The latter is a conserved quantity,
since there is no flow through the boundaries. The unit of
velocity is derived from the basic units as $[u]=[x]/[t]=b \varOmega$, and the
unit of acceleration is $[a]=[x]/[t]^2=b \varOmega^2$. For clarity we will
often write the units out explicitly.

\subsection{Global solutions}
\label{s_globalsolutions}

The velocity field given by \Eqss{uxgood}{uzgood} is not a global solution,
since the velocity field is discontinuous when crossing the vortex boundary to
the surrounding Kepler flow, where the only velocity component is
$u_y = -q \varOmega x$. By looking at the $y$-velocity of the flow,
\EQ
  u_y = - \frac{1}{\epsilon} \alpha \varOmega x = -
  \frac{\sqrt{3}}{\sqrt{1-\epsilon^2}} \varOmega x \, ,
\EN
it is apparent that, regardless of the choice of $\epsilon$, the tangential part
of the vortex flow will always be faster than the Kepler flow at any position
in $x$. It seems that for the vortices considered here there is no way of
producing a linear velocity field that can make a gradual transition from the
vortex to the surrounding Kepler flow.
Our initial conditions do therefore not satisfy our equations at the interface
between vortex and exterior.
For this reason we must expect to see an evolution away from our initial state.

\section{Dust-trapping mechanism}
\label{s_dusttrapping}

\subsection{Forces driving the vortex}

The fact that anticyclonic vortices suck in dust particles can be explained by
looking at the forces involved in the rotation. The rotation of gas is
maintained because the Coriolis force $\vec{F}_{\rm Cor}$, the tidal force
$\vec{F}_{\rm g+c}$ and the pressure gradient force $\vec{F}_{\rm p}$ add up to
a resulting force $\vec{F}_{\rm cen}$ directed towards the centre of the
coordinate system. The symbol $\vec{F}$ is here used for the force per unit
mass.

\begin{figure}[t!]
  \begin{center}
    \epsfig{file=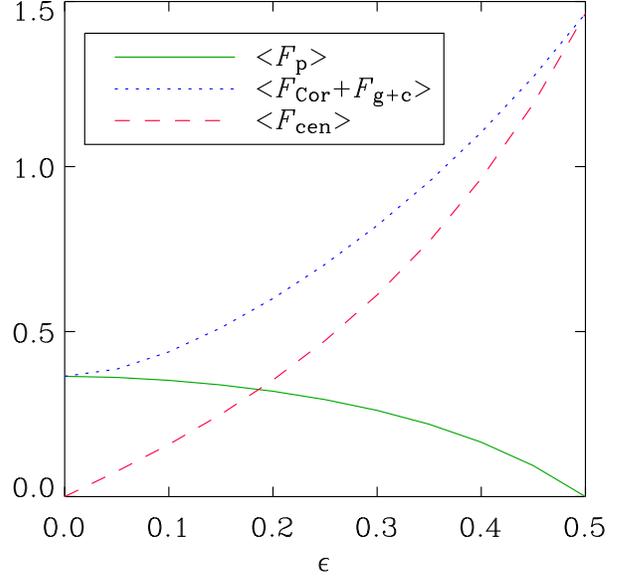}
  \end{center}
  \caption[]{Forces affecting the gas. Average magnitudes of pressure gradient
  force $F_{\rm p}$, Coriolis force $F_{\rm Cor}$ plus tidal force $F_{\rm g+c}$
  and resulting centre-directed force $F_{\rm cen}$ as functions of aspect
  ratio $\epsilon$. Averages are taken over the vortex boundary. The force unit
  is $b \varOmega^2$.}
  \label{f_epsforces}
\end{figure}

\begin{figure*}[t!]
  \begin{center}
    \epsfig{file=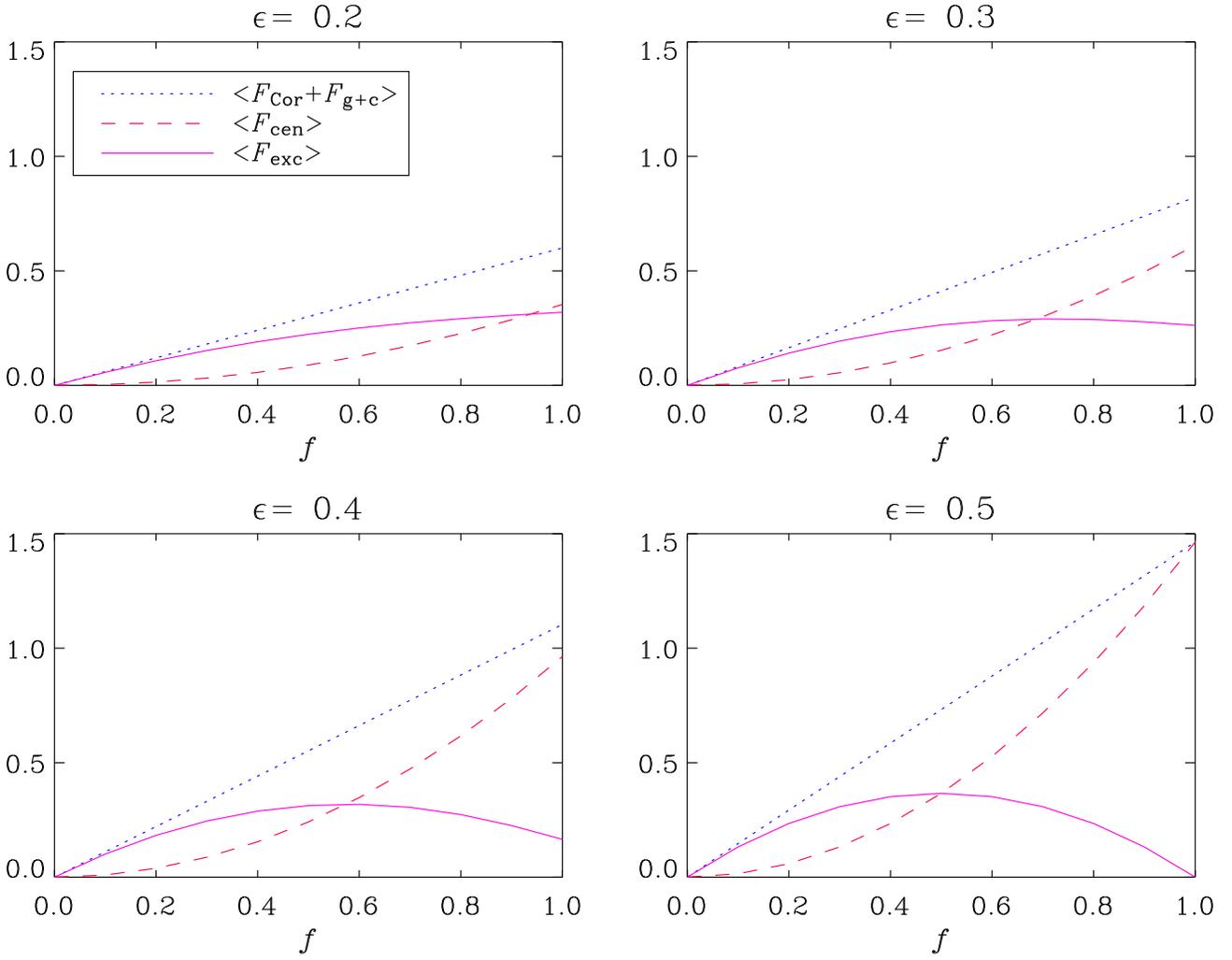}
  \end{center}
  \caption[]{Forces affecting dust as a function of $f$ (dust speed as a
  fraction of gas speed). Average values of Coriolis force $F_{\rm Cor}$ plus
  tidal force $F_{\rm g+c}$, necessary centre-directed force to maintain
  rotation $F_{\rm cen}$ and excess force $F_{\rm exc}$ for different values of
  aspect ratio $\epsilon$.  The force unit is $b \varOmega^2$. Averages are
  taken over the vortex boundary. The forces do not have the same direction, so
  the two vectors have been added to get the magnitude of the excess force. The excess
  force converges to the missing pressure gradient force for $f=1$ in all
  cases.}
  \label{f_forces}
\end{figure*}

The pressure gradient force is calculated from \Eq{hgood} (we consider $z=0$
since the pressure gradient in the $z$-direction is completely balanced by the
vertical gravity component),
\EQA
  \vec{F}_{\rm p} &=& - \nab h =
  \pmatrix{\delta^2 \varOmega^2 x \cr \delta^2 \varOmega^2 \epsilon^2 y \cr 0}
  = \delta^2 \varOmega^2 \pmatrix{x \cr \epsilon^2 y \cr 0} \nonumber \\ 
  &=& \left( -\frac{3}{1-\epsilon^2}+\frac{2\sqrt{3}}{\sqrt{1-
  \epsilon^2}} \right) \varOmega^2 \pmatrix{x \cr \epsilon^2 y \cr 0} \, ,
\ENA
and the Coriolis force is
\EQA
  \vec{F}_{\rm Cor} &=& -2 \varOmega \hat{\vec{k}} \times \vec{u} = -2 \varOmega
  \pmatrix{0 \cr 0 \cr 1} \times \pmatrix{\epsilon \alpha \varOmega y \cr
  - \alpha \varOmega x/\epsilon \cr 0} \nonumber \\
  &=& -2 \varOmega
  \pmatrix{ \alpha \varOmega x/\epsilon \cr \epsilon \alpha \varOmega
  y \cr 0} = -\frac{2\alpha}{\epsilon}\varOmega^2 \pmatrix{x \cr \epsilon^2 y
  \cr 0} \nonumber \\
  &=& -\frac{2\sqrt{3}}{\sqrt{1-\epsilon^2}} \varOmega^2 \pmatrix{x \cr
  \epsilon^2 y \cr 0}\, .
\ENA
Note that both forces point perpendicular
to the vortex flow. The resulting force,
\EQA
  \vec{F}_{\rm res} &=& \vec{F}_{\rm p} + \vec{F}_{\rm Cor} + \vec{F}_{\rm g+c}
  \nonumber \\
  &=& -\frac{3}{1-\epsilon^2} \varOmega^2 \pmatrix{x \cr \epsilon^2 y \cr 0} +
  \pmatrix{3 \varOmega^2 x \cr 0 \cr 0} \nonumber \\
  &=& -\frac{3}{1-\epsilon^2} \varOmega^2 \pmatrix{x \cr \epsilon^2 y \cr 0}
  -\frac{3}{1-\epsilon^2} \varOmega^2 \pmatrix{(\epsilon^2-1) x \cr 0 \cr 0}
  \nonumber \\
  &=& -\frac{3}{1-\epsilon^2} \varOmega^2 \pmatrix{\epsilon^2 x \cr \epsilon^2
  y \cr 0} = -\frac{3\epsilon^2}{1-\epsilon^2}\varOmega^2\pmatrix{x \cr y \cr0} 
  \, ,
\ENA
points always towards the centre and thus produces the rotating motion.
\Fig{f_epsforces} is a plot of the average values over the vortex boundary of
the pressure gradient force, Coriolis force plus tidal force and resulting
centre-directed force as a function of $\epsilon$. The force magnitudes have
the unit of $b \varOmega^2$. As the aspect ratio decreases, the pressure
gradient force becomes relatively more important. At $\epsilon = 0.5$ the
vortex is in equilibrium without any need for a pressure gradients in the
$x$-$y$ plane. This pressure-less vortex is exactly the epicyclic vortex
considered by Barge \& Sommeria (\cite{barge+sommeria95}). 

\subsection{Forces on the dust}

Dust, initially at rest, is accelerated by drag to follow gas around
the vortex, but in the beginning with a much smaller velocity. Moving at a
fraction $f$ of the gas speed relative to the main shear flow,
$\tilde{\vec{u}}_{\rm d} = f \tilde{\vec{u}}$,
we find for $ \vec{u}_{\rm d} = \tilde{\vec{u}}_{\rm d} +\vec{u}^{(0)}$
the expression
$ \vec{u}_{\rm d} = f\vec{u} + (1-f) \vec{u}^{(0)}$, so dust is subjected to
the Coriolis force
\EQ
  \vec{F}_{\rm Cor} = -\frac{2\sqrt{3}}{\sqrt{1-\epsilon^2}} f \varOmega^2 
  \pmatrix{x \cr \epsilon^2 y \cr 0} - (1-f) \pmatrix{3 \varOmega^2 x \cr 0
  \cr 0}\, . 
\EN
The centre-directed force needed to keep dust spinning around
the vortex is proportional to velocity squared, and is therefore reduced by a
factor of $f^2$ compared to the force needed to keep the gas rotating,
\EQ
  \vec{F}_{\rm cen} = -\frac{3 f^2 \epsilon^2}{1- \epsilon^2}
  \varOmega^2\pmatrix{x \cr y \cr 0} \, . 
\EN
The excess force on the dust is then $\vec{F}_{\rm exc} = \vec{F}_{\rm Cor} +
\vec{F}_{\rm g+c} - \vec{F}_{\rm cen}$. In \Fig{f_forces} Coriolis force
plus tidal force, centre-directed force needed to maintain the same radius of
rotation as the gas and excess force working on the dust are shown as a
function of $f$ and for different values of the aspect ratio $\epsilon$. For all value of
$\epsilon$ the Coriolis force grows much faster with $f$ than the required
centre-directed force. This leads to an excess force directed inwards, which
causes dust to spiral inwards. Similar explanations for the dust-trapping
mechanism of vortices are given in Tanga et al.\ (\cite{tanga+etal96}) and
Chavanis (\cite{chavanis00}). Already at a few percent of the gas speed, the
excess force on the dust is significant. Cyclonic vortices have an outwards
directed Coriolis force. They will therefore not be able to concentrate solid
particles, but will rather expel them.

The centre-directed force catches up with the Coriolis force at $f=1.0$ for
$\epsilon=0.5$, but for higher $\epsilon$ there is still an excess force on the
dust at $f=1.0$ of the same magnitude (but opposite direction) as the pressure
gradient force affecting the gas. The reason for this is that dust particles
are not subjected to the pressure gradient force of gas, so even if dust particles
were light enough to quickly become accelerated to match gas velocity, they
will feel an extra force inward and thus begin to spiral inwards.
The terminal velocity
that dust obtains perpendicular to the gas flow, $\vec{u}_{{\rm d}
\perp}^{\rm (ter)}$, is a result of a balance between excess inwards force and
drag force,
\EQ
  0 = - \delta^2 \varOmega^2 \pmatrix{x \cr \epsilon^2 y \cr 0} -
  \frac{1}{\tau_{\rm s}} \vec{u}_{{\rm d}\perp}^{\rm (ter)} \, ,
  \label{udriftbalance}
\EN
which means that
\EQ
  \vec{u}_{{\rm d}\perp}^{\rm (ter)} =
  - \tau_{\rm s} \delta^2 \varOmega^2 \pmatrix{x \cr \epsilon^2 y \cr 0} \, ,
  \label{udriftmax}
\EN
in analogy to the vertical settling of dust towards the mid-plane due to
gravity. The time scale for this is quite long,
\EQ
  t \sim \frac{1}{\tau_{\rm s} \varOmega^2} \, ,
\EN
considering that the stopping time must be short for the dust to match the
velocity of the gas, but if vortices are indeed long-lived, it could prove an
important mechanism for trapping dust particles of short stopping time inside vortices of low 
aspect ratio.

The aspect ratio also has an influence on dust-trapping at low $f$. It is
evident from \Fig{f_forces} that the excess force grows faster with $f$ and
reaches a higher maximum for high $\epsilon$.

\section{Three-dimensional models}
\label{s_3dmodels}

The vortex solution of GNG has zero enthalpy
outside the vortex, corresponding to zero density through the relation
\EQ
  \rho = \rho_0 \left[ (\gamma-1) \frac{h}{c_{\rm s0}^2}
  {\rm e}^{\gamma s/c_{\rm p}} \right]^{1/(\gamma-1)} \, .
  \label{enthtoden}
\EN
This gives rise to a potential problem in
three dimensions: hydrostatic equilibrium in the $z$-direction requires
\EQ 
  - \frac{\partial \varPhi}{\partial z} - \frac{1}{\rho} \frac{\partial  
  P}{\partial z} = 0 \, ,
  \label{hydequi}
\EN
where
\EQ 
  \varPhi=\frac{1}{2} \varOmega^2 z^2
\EN
is the vertical gravitational potential. This means that the pressure must fall
off in the vertical direction, but it must never become negative, so embedding
the vortex in a disc of finite density and constant entropy is not possible in the vertical direction.

We use two different ways of modelling the vortices in three dimensions:
hot corona and cylindrical vortex.

\subsection{Hot corona}

Hydrostatic equilibrium in the $z$-direction can be obtained by embedding the
vortex in a tenuous gas of high temperature, i.e.\ a hot corona.
In terms of specific
enthalpy, \Eq{hydequi} can be written as \EQ \frac{\partial \varPhi}{\partial z}
= \frac{h}{c_\mathrm{p}} \frac{\partial s}{\partial z} - \frac{\partial
h}{\partial z} \, .  \EN For a given entropy distribution it is then possible
to construct a continuous enthalpy so that hydrostatic equilibrium is obtained.
This method is also used by von Rekowski et al.\ (\cite{vReko_etal03}).

The hot corona simulations are done in a box of size $(L_x,L_y,L_z)/b=(4,32,4)$. 

\subsection{Cylindrical vortex}

Here we extend the vortex to the entire $z$-length of
the box. The vortex is then no longer ellipsoidal, but rather cylindrical with
an ellipse-shaped cross section. Hydrostatic equilibrium is obtained without
use of entropy through
\EQ
  \frac{\partial h}{\partial z} = -\frac{\partial \varPhi}{\partial z} =
  -\varOmega^2 z,
\EN
so $h(z) = -\frac{1}{2} \varOmega^2 z^2 +h_0$,
where $h_0$ must be greater than $\frac{1}{2} \varOmega^2 z_{\rm top}^2$ to
avoid negative enthalpy anywhere.
The cylindrical vortex simulations are done in a box of size
$(L_x,L_y,L_z)/b=(4,32,0.4)$. The reason for using a shallower box for the
cylindrical vortex is to allow for a smaller $h_0$ and thus to have a large density ratio between the vortex and its
surroundings (see \Sec{s_lifetime}). The minimum enthalpy in the mid-plane is $\frac{1}{2} \varOmega^2
z_{\rm top}^2$, so lowering $z_{\rm top}$ and adopting the minimal enthalpy
possible thus leads to a larger density
ratio.

In both cases we let dust start with zero initial velocity. Dust
density at height $z$ is initialised to be a fraction $r=0.01$ of gas
density  (see Natta et
al.\ \cite{natta+etal00}) far away from the vortex at the same height. Although the actual value
of $r$ does not directly influence dust dynamics, it does influence the
amount of back-reaction from dust upon gas.

\begin{figure}[t!]
  \begin{center}
    \includegraphics[]{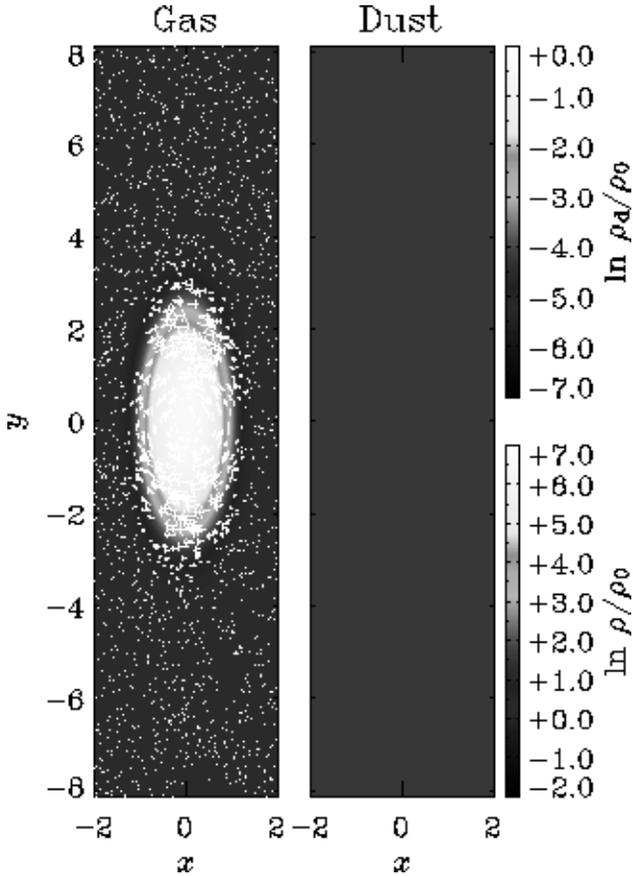}
    \caption[]{
      \label{f_init_xy}
      The initial condition in the mid-plane for the hot corona vortex:
      gas density and gas velocity
      (left) and the corresponding dust density and velocity (right). The
      velocity field does not seem to follow the contours of the vortex, but
      this is because the Kepler velocity has been subtracted.}
  \end{center}
\end{figure}

\section{Numerical method and boundary conditions}
\label{s_numericalmethod}

We simulate the motion of gas and dust in a coordinate frame corotating
at the local Keplerian speed. We use the Pencil-Code\footnote{The code is
available at \\ \url{http://www.nordita.dk/data/brandenb/pencil-code}}
(Brandenburg \& Dobler \cite{brandenb+dobler02}) which uses third order
Runge-Kutta time stepping and a sixth order finite-difference scheme in space
and employs central finite differences, so the extra cost of recentering a
large number of variables between staggered meshes each time step is avoided.
The code solves the non-conservative form of the equations
(see Brandenburg \cite{brandenb03} for details).

Periodic boundary conditions for velocities, density and specific entropy are used in
the $x$- and $y$-directions. The periodic boundary condition in $x$ is
appropriately sheared in the shearing sheet approximation. In the $z$-direction
we use stress-free boundary conditions, i.e.\
\EQ
u_{x,z}=u_{y,z}=u_z=s=
u_{{\rm d}x,z}=u_{{\rm d}y,z}=u_{{\rm d}z}=0,
\EN
where commas denote partial differentiation.
Dust and gas densities on the boundaries are fully determined
by the equations, but in practice we need to specify values in the
ghost zones which is accomplished by extrapolation.
We use a resolution of $(n_x,n_y,n_z)=(128,256,128)$ grid points.

\section{Results}
\label{s_results}

\subsection{Initial conditions and choice of stopping times}

\begin{figure}[t!]
  \begin{center}
    \includegraphics[]{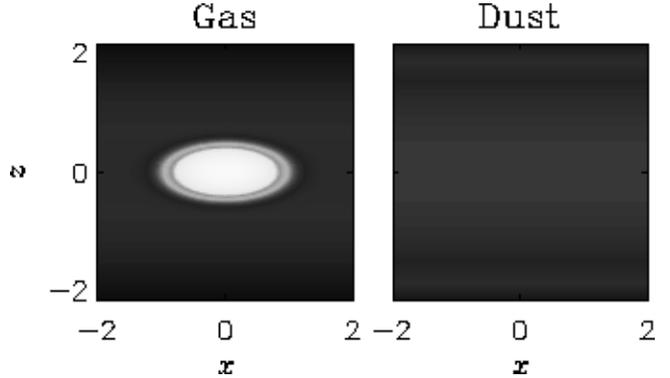}
    \caption[]{
      \label{f_init_xz}
      The initial condition in the vertical for the hot corona vortex:
      gas density
      (left) and the corresponding dust density (right). Dust density as a
      function of $z$ is
      initially 0.01 times gas density in the same height.
      The density colour scale is the same as in \Fig{f_init_xy}.}
  \end{center}
\end{figure}
The initial condition for gas and dust in the hot corona model is plotted in
\Fig{f_init_xy} (mid-plane
, only one fourth of the $y$-length of the box is shown) and
\Fig{f_init_xz} ($x$-$z$ plane). We choose an aspect ratio of
$\epsilon=0.4$ since we expect dust-trapping to be most efficient for vortices
of high aspect ratio. The transition from vortex to surroundings has been
smeared out both in density and velocity field. We have experienced that the
simulations run better when abrupt transitions are smoothed out. The dust is
initially at rest. In all plots of velocity fields we use the
Kepler-subtracted velocities $\tilde{\vec{u}}$ and $\tilde{\vec{u}}_{\rm d}$.

\begin{figure*}[t!]
  \begin{center}
    \epsfig{file=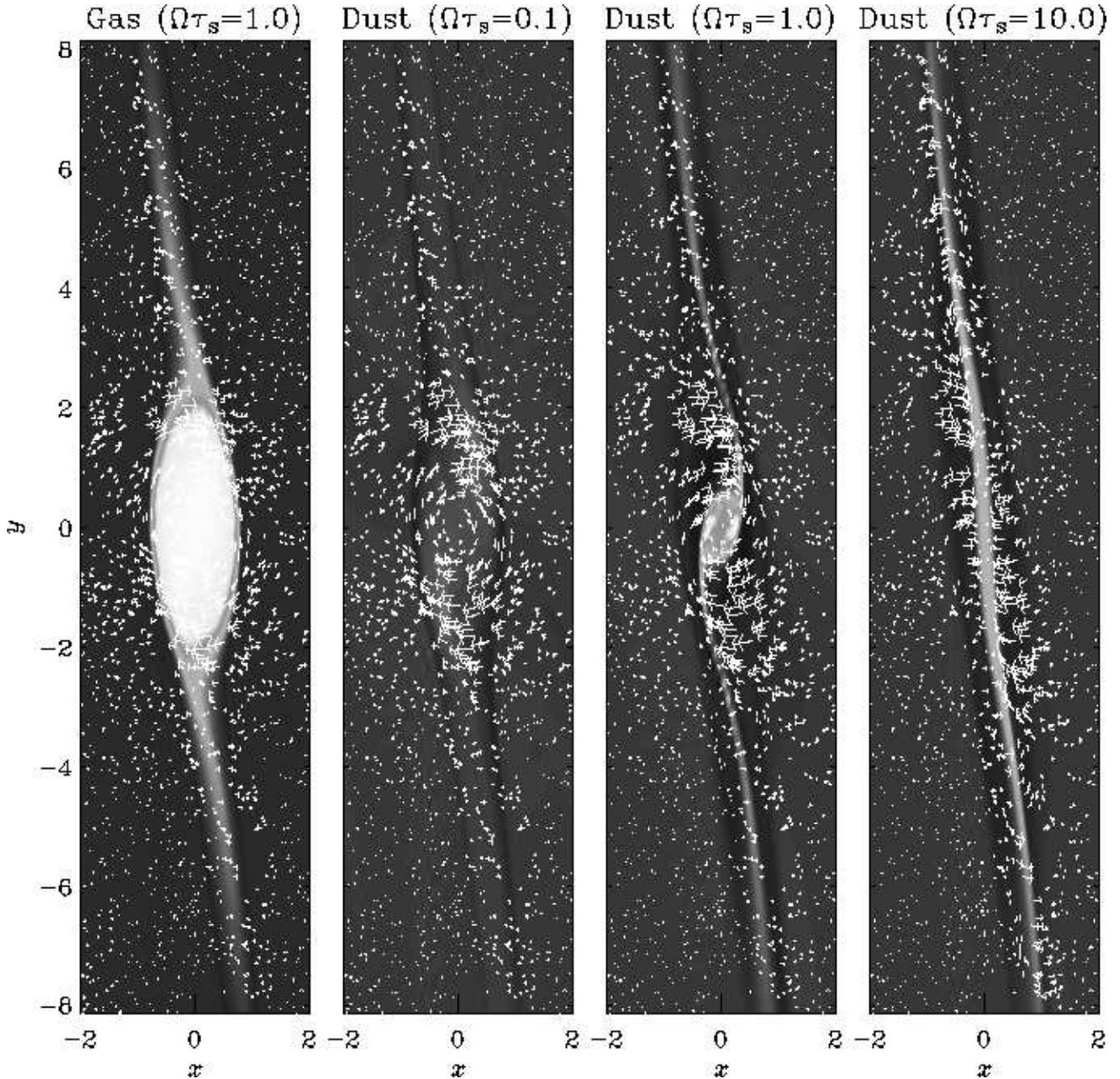}
  \end{center}
  \caption[]{Gas and dust in the mid-plane for the hot corona after one orbit.
  Gas density and velocity field for the three values of
  $\varOmega\tau_{\rm s}$ are indistinguishable, so only $\varOmega\tau_{\rm
  s}=1.0$ is shown. The other three plots show dust density and dust
  velocity field after one orbit for $\varOmega\tau_{\rm s}=0.1$,
  $\varOmega\tau_{\rm s}=1.0$ and $\varOmega\tau_{\rm s}=10.0$ respectively.
  The dust velocity field of $\varOmega\tau_{\rm s}=0.1$ is very similar to
  that of the gas, due to the short stopping time. For $\varOmega\tau_{\rm
  s}=1.0$ there is a strong convergence towards the interior of the vortex,
  whereas for $\varOmega\tau_{\rm s}=10.0$ only a slight density increase
  in a narrow region that extends from the vortex along the shear is seen.
  The density colour scales are the same as in \Fig{f_init_xy}.}
  \label{f_panel_xy}
\end{figure*}

\begin{figure*}[t!]
  \begin{center}
    \epsfig{file=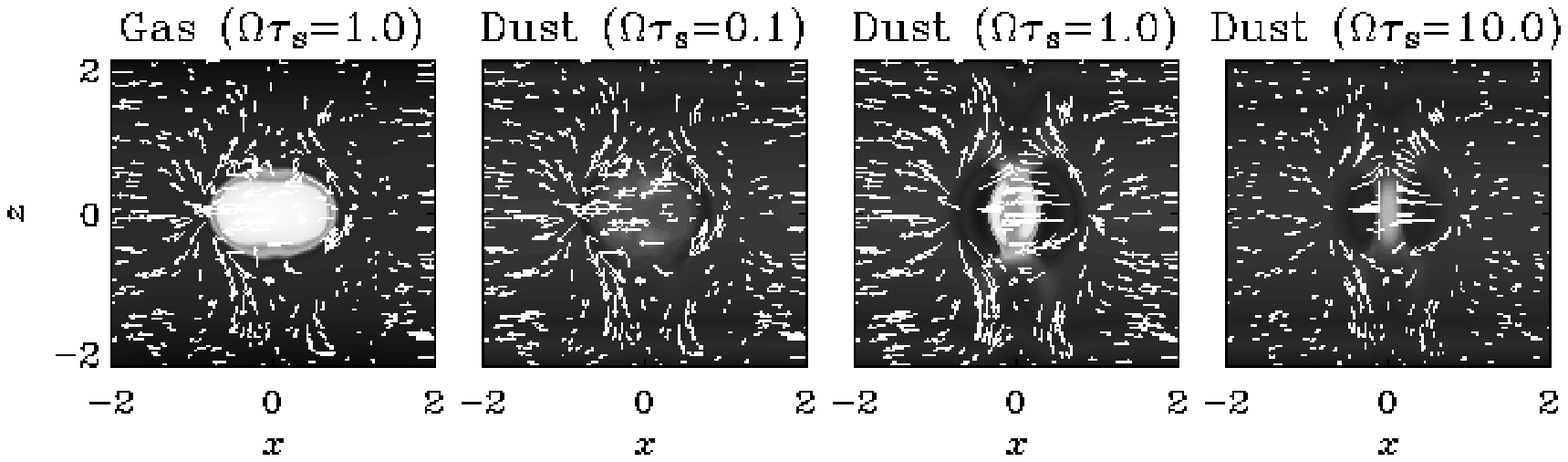}
  \end{center}
  \caption[]{Gas and dust in the $x$-$z$ plane for the hot corona case after one orbit.
  Gas density and velocity field for the three values of
  $\varOmega\tau_{\rm s}$ are indistinguishable, so only $\varOmega\tau_{\rm
  s}=1.0$ is shown. The other three plots show dust density and dust
  velocity field after one orbit for $\varOmega\tau_{\rm s}=0.1$,
  $\varOmega\tau_{\rm s}=1.0$ and $\varOmega\tau_{\rm s}=10.0$ respectively.
  For $\varOmega \tau_{\rm s}=0.1$ dust is so strongly coupled to gas that the velocity
  fields are very similar, whereas $\varOmega \tau_{\rm s}=1.0$ shows strong
  convergence of dust in the interior of the vortex. For $\varOmega \tau_{\rm
  s}=10.0$ dust is not as affected by gas motion, although strong motions
    move dust around the vortex and upwards. Velocities are exaggerated about 10
  times compared to \Fig{f_panel_xy}. The density colour scales are the same as
  in \Fig{f_init_xy}.}
  \label{f_panel_xz}
\end{figure*}

We have run hot corona simulations for three different values of the stopping
time: $\varOmega\tau_{\rm s}=0.1$ (dust completely coupled to gas), $\varOmega\tau_{\rm s}=1.0$
(dust semi-coupled to gas) and $\varOmega\tau_{\rm s}=10.0$ (dust almost not coupled to
gas). This corresponds to rocks of sizes respectively $a_{\rm
s}=10~\rm{cm}$, $a_{\rm s}=100~\rm{cm}$ and $a_{\rm s}=1000~\rm{cm}$,
respectively. For the minimum-mass nebula of Cuzzi et al.\
(\cite{cuzzi+etal93}) the mean free path is $\lambda = (r/{\rm AU})^{11/4} \,
{\rm cm}$. This
means that we must, strictly speaking, go beyond $r = 10 \, {\rm AU}$ for the
largest particles to be in the Epstein regime. For $a_{\rm s}=10~\rm{cm}$ and
$a_{\rm s}=100~\rm{cm}$ the Epstein drag law is valid already before 
$r = 5 \, {\rm AU}$. The range of dust radii considered is expected for
the largest agglomorations of sticking dust particles after having reached the mid-plane, according to
Weidenschilling \& Cuzzi (\cite{weidenschilling+cuzzi93}) who model how
dust particles falling towards the mid-plane stick to each other in a turbulent protoplanetary disc.
Larger sizes are prohibited by the turbulence in the disc.

The cylindrical vortex was only run for $\varOmega \tau_{\rm s}=1.0$, since we are mostly
interested in whether it is a valid 3D vortex solution.
We follow the evolution of gas and dust
for a full Kepler orbit $t_{\rm max} = 2 \pi / \varOmega$ (which is not quite
a full rotation of the vortex, see \Fig{f_angvel}). 

\subsection{Viscosity parameters}

For the hot corona model we are able to run for a whole orbit with a viscosity of $\nu =
\nu_{\rm d} = 2 \cdot 10^{-4}$
and a shock viscosity of $c_{\rm shock} = 2$.
This corresponds to a mesh Reynolds number of
\EQ
  \mbox{Re}_{\rm mesh} = \max(|\tilde{\vec{u}}|) \max(\delta x, \delta y,
  \delta z) / \nu \approx 625 \, .
\EN
Thus, since $\mbox{Re}_{\rm mesh}$ is rather large, the viscosity $\nu$
is almost everywhere unimportant, and most dissipations occurs through
the shock viscosity (not included in the expression for
$\mbox{Re}_{\rm mesh}$), but this affects only convergent flow regions
and not the vortical parts of the flow.
This can be seen from the following consideration.

The Laplacian of
$\tilde{\vec{u}}$ can be written $\nab^2 \tilde{\vec{u}} = - \nab \times \nab
\times \tilde{\vec{u}} + \nab \nab \cdot \tilde{\vec{u}}$. Using this we can
rewrite the expression for the viscous force as
\EQ
  \frac{\vec{F}_{\rm visc}}{\rho} = \nu \left(-\nab \times
  \tilde{\vec{\omega}} + 2 \vec{\mathsf{S}} \cdot \nab \ln \rho \right) +
  ({\textstyle\frac{4}{3}} \nu + \zeta) \nab \nab \cdot \tilde{\vec{u}} \, ,
  \label{viscosity2}
\EN
where $\tilde{\vec{\omega}}=\nab\times\tilde{\vec{u}}$ is the vorticity of the flow.
This shows that the kinematic shear
viscosity works as a drain for vorticity in regions where the
vorticity changes, whereas shock viscosity only affects regions of
convergence in the velocity field.

At the boundary
between vortex and surroundings the vorticity changes abruptly, 
and we have experienced vortices dissolving rapidly
when using too high a viscosity. We stress that this is a purely
numerical issue, and that real physical viscosity is many orders of
magnitude lower than the viscosity we use here.

For the cylindrical vortex we had to use a
viscosity of $\nu = \nu_{\rm d} = 10^{-2}$ and a shock viscosity of $c_{\rm
shock} = 4$. The reason for these higher viscosities is that the cylindrical
vortex does not have as large a density ratio as the hot corona vortex,
so it takes more viscosity to keep the vortex intact on the boundary.

\subsection{Lifetime of vortices}
\label{s_lifetime}

When using a realistic disc background density we have experienced the vortices
breaking up at the local sound speed at the vortex boundary. This we believe is caused by
the term $\nab \cdot \tilde{\vec{u}}$ in the continuity equation \eq{contgas}. Although
the initial condition has $\nab \cdot \tilde{\vec{u}} = 0$ analytically, this is not
necessarily valid numerically. Calculating the spatial derivatives of the
velocity field on a Cartesian coordinate grid yields $\nab \cdot \tilde{\vec{u}}
\neq 0$ over a few mesh-points (due to the order of the numerical
derivatives) around the vortex boundary. The result is the depletion of density at
the NW and SE corners of the vortex, and increase of density at the NE and SW
corners. We find that we can alleviate this problem by having a large density
contrast between the vortex and its exterior [as also suggested by the GNG 
vortex solution, cf.\ \Eq{hgood}]. We also find that shock viscosity helps
to reduce this problem.

For the hot corona, we plot in \Fig{f_panel_xy}
gas density, gas velocity, dust density and dust velocity in the mid-plane after
one orbit. The gas
configuration is indistinguishable between different values of $\varOmega \tau_{\rm
s}$, indicating that there is very little back-reaction on the gas, so gas
density and velocity field are only shown for $\varOmega \tau_{\rm s}=1.0$.

The vortex is evidently still in place, although the
outer parts have been sheared away to form long tails in the direction of the
shear. In \Fig{f_panel_xz}
gas density and dust density in the $x$-$z$ plane are plotted.
Again we see that the vortex is still in place after one orbit. We conclude
from this that the hot corona vortex solution is valid even in the presence of
shear.

\begin{figure}[t!]
  \begin{center}
    \epsfig{file=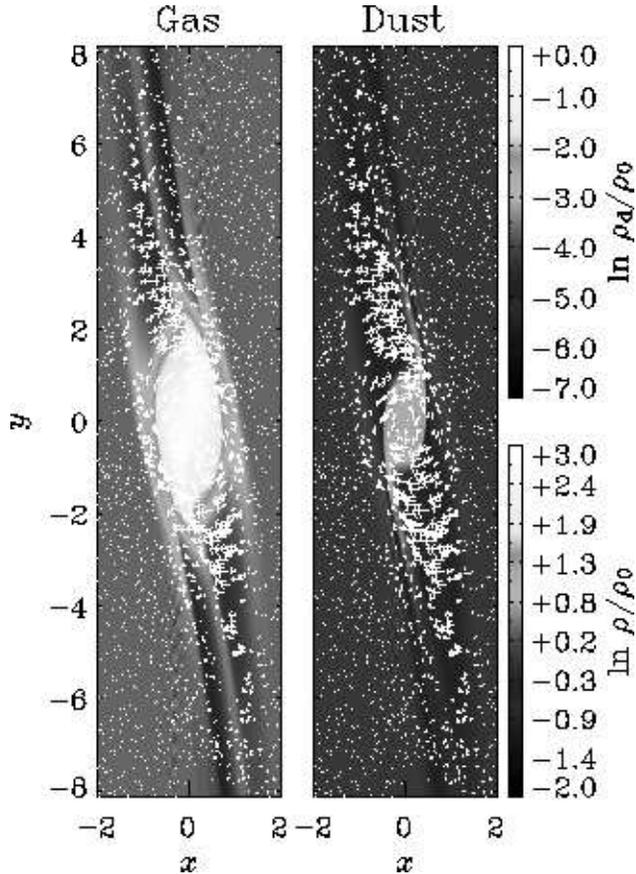}
  \end{center}
  \caption[]{Gas and dust in the mid-plane for the cylindrical vortex after one
  orbit, $\varOmega\tau_{\rm s}=1.0$. The vortex is close to breaking up: two
  arcs of matter are being expelled at the NE and SW corners. Note that the
  density contrast between the vortex and its surroundings is much smaller than
  for the hot corona vortex, and that the gas density colour scale is different
  from the hot corona's.}
  \label{f_taus1.0cyl_xy}
\end{figure}

The status of the cylindrical vortex after one orbit is shown in
\Fig{f_taus1.0cyl_xy}. In the mid-plane it seems to be more
disrupted than the hot corona vortex, possibly due to the smaller density ratio
between the vortex and its surroundings. A vertical cut (not shown) reveals
that the
original strati\-fication of the vortex is almost intact, which implies that the
cylindrical vortex model is also a valid one.

\begin{figure}[t]
  \begin{center}
    \epsfig{file=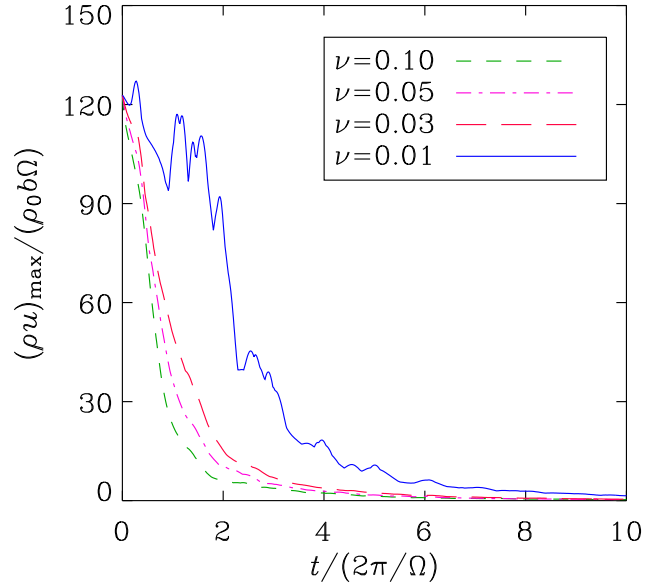}
  \end{center}
  \caption[]{Maximum mass flux in the box as a function of time (measured in
  orbits) for different values of the viscosity $\nu$. A low viscosity
  obviously increases the lifetime of a vortex. On the other hand a low
  viscosity also leads
  to more chaotic development in $(\rho u)_{\rm max}$, 
  indicating that the mesh
  Reynolds number may become too high.}
  \label{f_viscosity}
\end{figure}

To test the lifetimes of vortices beyond one orbit we have run low-resolution
simulations ($64 \times 128 \times 64$) for different values of the viscosity
$\nu$. Since the major effect of viscosity takes place at the vortex boundary
where the velocity (and thus the mass flux) is highest, we plot the maximum mass flux $(\rho
u)_{\rm max}$ in the box as a
function of time measured in orbits in \Fig{f_viscosity}. The lifetime is
obviously very dependent on the viscosity, with lower viscosities leading to
longer-living vortices. A low viscosity, on the other hand, also leads to more
chaotic development in maximum mass flux, probably due to too high a mesh
Reynolds number.

\subsection{Evolution of dust density}
\label{s_dustdensity}

Also shown in \Figs{f_panel_xy}{f_panel_xz} is the evolution of dust density.
As expected, the most
efficient dust-trapping occurs when dust and gas are semi-coupled with
$\varOmega\tau_{\rm s}=1.0$, where the
increase in dust density inside the vortex is from an initial $\ln \rho_{\rm
d}/\rho_0=-4.18$ to a maximum of $\ln \rho_{\rm d}/\rho_0=-1.34$ after one orbit, a 17 times
increase in density. A depletion of dust in the outer parts of the vortex is
also seen. We believe this is where the trapped dust has been taken from. There
is a strong convergence in the dust velocity field both in the mid-plane and in
the $x$-$z$ plane.

\begin{figure}[t]
  \begin{center}
    \epsfig{file=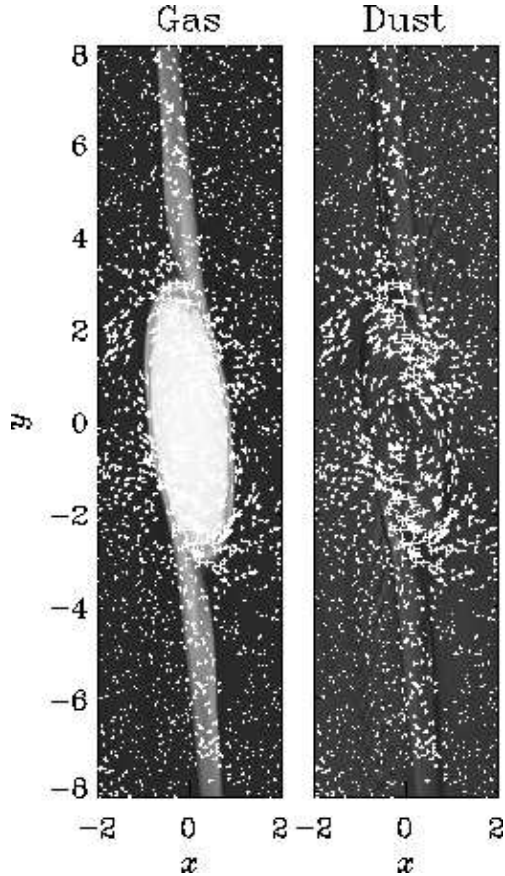}
  \end{center}
  \caption[]{Gas and dust density and velocity fields after 1.5 orbits for
  $\varOmega\tau_{\rm s}=0.1$. Comparing with \Fig{f_panel_xy} (plot number two
  from left) there is no apparent widening of the dust-depleted region on the
  vortex boundary, nor has the dust density increased inside the vortex. The
  colour scale is the same is in \Fig{f_init_xy}.}
  \label{f_settling}
\end{figure}

For $\varOmega\tau_{\rm s}=0.1$ dust is accelerated to match the gas velocity
very quickly (compared to an orbit) and only a very modest increase in dust
density is seen inside the vortex. This could be the slow settling of dust that
was discussed in \Sec{s_dusttrapping}, which should here happen on a time scale
of about $10/\varOmega$. The velocity field of \Fig{f_panel_xz} does however
not show any convergence inside the vortex. But according to \Eq{udriftmax} the
inwards drift could happen with a velocity of only $u^{\rm(ter)}/(b
\varOmega) \approx \tau_{\rm s} \varOmega \delta^2 \approx 0.02$ for
$\varOmega\tau_{\rm s}=0.1$ and $\epsilon=0.4$, a contribution in peril of
drowning completely in the erratic motions present after one orbit,
although it could still be present underneath it. A narrow region of dust
depletion is apparent around the edge of the vortex. This may be caused by the
slow inwards drift of dust (its radius of $\sim 0.1$ is comparable to the
expected $0.02\cdot2 \pi = 0.13$), although it is not clear why the $\varOmega
\tau_{\rm s}=10.0$ vortex should have a region of similar depletion. In
\Fig{f_settling} we show gas and dust configurations of the
$\varOmega\tau_{\rm s}=0.1$ vortex after 1.5 orbits. The dust-depleted region
on the vortex boundary has obviously not become wider, nor has the dust density
inside the vortex increased. Gas density has changed a lot. The vortex now
has a pronounced tilt in the NW-SE direction, and the shear tails have grown
more massive in density. The reason why we do not see any dust-trapping may
then very well be that vortex dynamics completely dominates over the slow
inwards drift of dust, rendering the mechanism for trapping short stopping time
dust particles useless.

For a stopping time parameter of $\varOmega\tau_{\rm s}=10.0$ dust is so unaffected by
gas that only a slight density increase is seen. The increase has occurred
in a narrow region that extends from the vortex along the shear. This may be
the dust-trapping mechanism 
that here takes place so slowly that dust is sheared away faster
than the vortex can trap it.

\subsection{Back-reaction on gas}

\begin{figure}[t]
  \begin{center}
    \epsfig{file=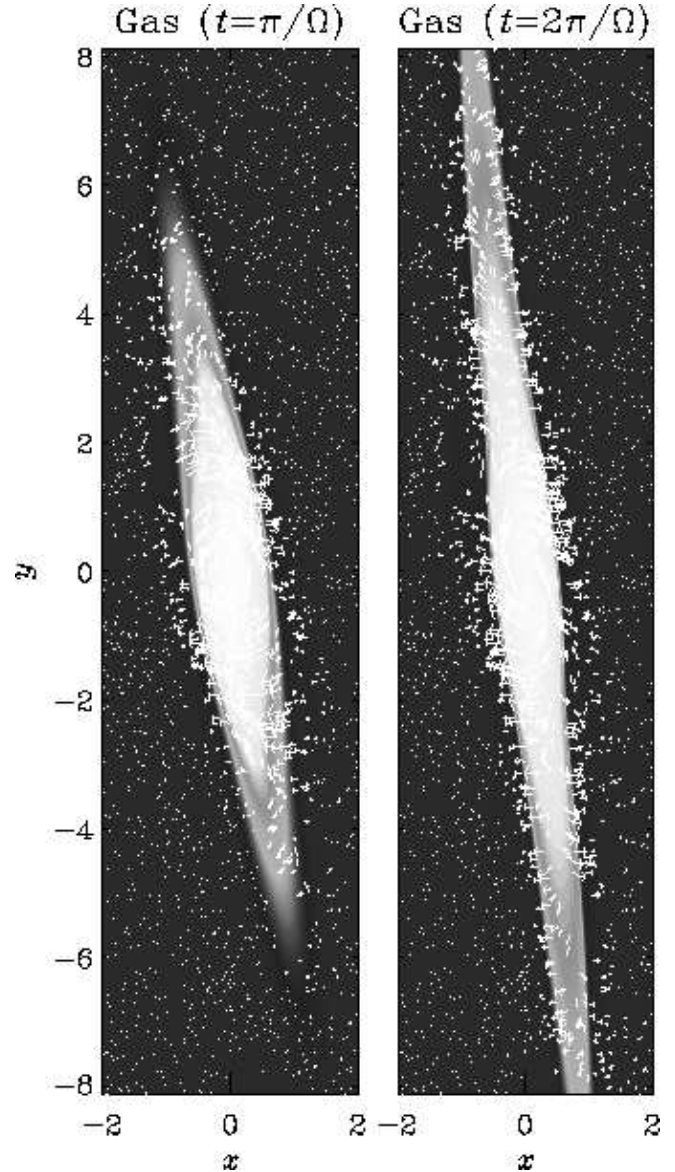}
  \end{center}
  \caption[]{Gas configuration after half an orbit (left) and after a whole
  orbit (right) for a dust-to-gas ratio of the order of $1$ inside the vortex
  ($100$ in the disc). Velocities are reduced by around a factor of 5
  compared to the initial condition. Gas is forced to follow the converging
  dust, and that leads to the destruction of the vortex. The
  colour scale is the same is in \Fig{f_init_xy}.}
  \label{f_backreaction}
\end{figure}

One very important issue regarding vortices is how they are affected by the
dust they trap. Vertical dust-settling at the same time increases dust
density in the mid-plane, and at one point dust density in the vortex is
expected to become comparable to gas density. Then the back-reaction on gas
becomes important in the dynamics of the gas, \Eq{tstopgd}.

As mentioned in \Sec{s_lifetime} the reason why gas develops similarly in our
simulations for all values of stopping time is that the back-reaction on the
gas from dust is negligible. This is caused by two things: a dust-to-gas
ratio of $0.01$ is not very high, and furthermore this ratio only applies in
the disc, whereas we know that gas density in the vortex is very much
higher, giving an even lower dust-to-gas ratio there. The back-reaction would
eventually become much bigger if we allowed for vertical settling of dust.

To test when the back-reaction from dust becomes important we have run the
$\varOmega \tau_{\rm s}=1.0$ and $\varOmega \tau_{\rm s}=10.0$ vortices with
higher values of the dust-to-gas ratio. In \Fig{f_backreaction} we plot the gas
configuration of the $\varOmega \tau_{\rm s}=1.0$ vortex when the dust-to-gas
ratio is $100$. At this point dust density becomes comparable to gas density
inside the vortex. We see that the vortex is not only affected by dust drag,
but that it is even almost completely destroyed after one orbit. When dust
converges inside the vortex, it will drag gas along with it, in this way
destroying the vortex.

\begin{figure}[t]
  \begin{center}
    \epsfig{file=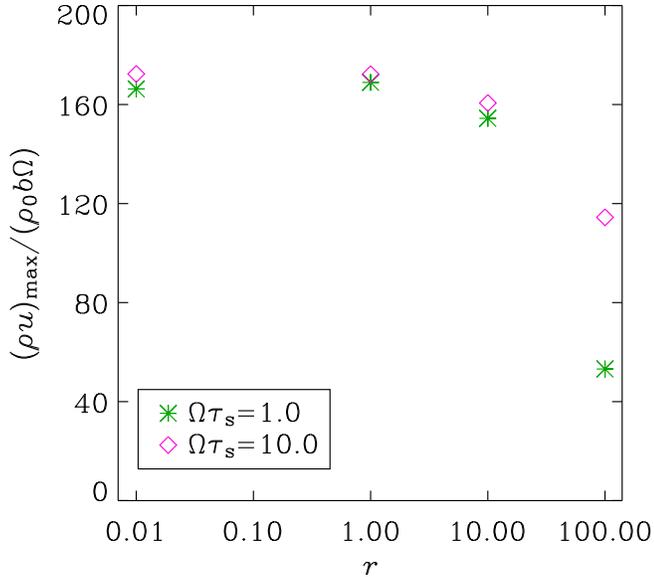}
  \end{center}
  \caption[]{The maximum mass flux in the box after one orbit
  when varying the dust-to-gas ratio from 0.01 to 100, the
  latter corresponding to a dust-to-gas ratio of unity inside the vortex.
  Points are shown for both $\varOmega \tau_{\rm s}=1.0$ and $\varOmega
  \tau_{\rm s}=100.0$. The best dust-trapper, $\varOmega \tau_{\rm s}=1$, is
  also the best at destroying the vortex, due to the trapped dust dragging gas
  along with it.}
  \label{f_backreac}
\end{figure}

The maximum mass flux in the box after one orbit as a function of dust-to-gas
ratio in the disc is shown in \Fig{f_backreac}. Already when the dust-to-gas
ratio in the disc is $10$ (corresponding to around $0.1$ in the vortex), the
back-reaction becomes important. The $\varOmega \tau_{\rm s}=1.0$ vortex is
best at destroying the vortex at all dust-to-gas ratios.

\section{Conclusions}
\label{s_conclusions}

In this paper we have explored the suggestion of Barge \& Sommeria
(\cite{barge+sommeria95}) of trapping dust particles in anticyclonic vortices.
In recent years the vortex theory has gained increasing popularity as a site
for planetesimal formation. 

To our knowledge this is the first time the dust-trapping mechanism has been
explored in three dimensions and with a freely developing gas velocity field
and density. We use two ways to model the vortices in three dimensions, and we have shown that both
models survive well for at least one orbit. It is conceivable that long life times are possible at yet higher resolution
when the viscosity can be smaller. We have also shown the
dust-trapping capabilities of vortices. For both vortex models dust-trapping is very
efficient when dust and gas are semi-coupled, whereas too little or too
much coupling gives
no significant dust-trapping. The most efficiently trapped solid particles
have a radius of 100 centimetres. This may be in some conflict with meteoritic
evidence, where chondrules (the building blocks of the most pristine
meteorites) are typically up to 1 millimetre in radius. But meteorites on
Earth are found to originate in the asteroid belt and should therefore 
not necessarily say anything about the make up of pristine material in the
outer Solar System.

Our two three-dimensional models both have a large density contrast between the vortex and the
surrounding disc. A model with a realistic disc density profile could be
obtained by using coordinates more natural to the vortex flow, such as
elliptical coordinates. This would also require a version of the shearing sheet
approximation adapted to new coordinates (for spherical coordinates there
exists the shearing disc approximation of
Klahr \& Bodenheimer \cite{klahr+bodenheimer03}). Alternatively, and probably better, a
global solution to the Euler and continuity equations with a gradual transition
from vortex to surrounding Kepler flow could be used. Such global solutions
have been found to the Euler equation alone (Chavanis \cite{chavanis00}, de la
Fuente Marcos \& Barge \cite{fuente+barge01}), but none to our knowledge
satisfy the continuity equation.

The role of vortices as high pressure regions seems hitherto unexplored. Short
stopping time solid 
particles forced to match gas speed around the vortex will feel an
additional pull inwards. The drift is similar to the vertical settling of dust
towards the mid-plane. Although this happens on quite a long time scale compared
to the conventional dust-trapping time, it seems a viable mechanism for
catching short stopping time 
dust particles, provided vortices live long enough. Unfortunately our
simulations do not show this mechanism at work, but rather that vortex and gas
dynamics completely dominates over this effect, even if it is present. 

It will be necessary to address the questions as to what causes
such vortices in the first place.
It seems plausible that anticyclonic vortices form as a relic
after the disc turbulence in the disc has died out
(Bracco et al.\ \cite{bracco+etal98}; \cite{bracco+etal99}).
This idea is particularly attractive, because the absence of turbulence
may be an important prerequisite for allowing dust to settle to
the mid-plane.
Other suggestions include baroclinic instabilities
(Klahr \& Bodenheimer, \cite{klahr+bodenheimer03}),
the anisotropic kinetic alpha-effect
(von Rekowski et al.\ \cite{vReko_etal99}),
and perhaps even the possibility of brown dwarf encounter
(Willerding \cite{willerding02}). The original work by GNG was inspired by
numerical results of the Papaloizou-Pringle instability, but this
instability is not available in a thin Kepler disc with $q=3/2$. Stability
analysis in GNG suggested that the vortices suffer from numerous linear
instabilities, so the survival of the vortices in 3D for at least one orbit
seems surprising.
In any case, it will be important to relax the need for implementing
vortices as initial conditions and rather try to get them
automatically under realistic conditions.

The photometric observations of the protostar KH15D has been interpreted by
Barge \& Viton (\cite{barge+viton03}) as a giant dusty vortex rotating at a
distance of 0.2 AU from the star and covering 120$^\circ$ of the orbit. Such a
large vortex is beyond the validity of the shearing sheet approximation
(where the curvilinearity of the disc is neglected) and requires a global
hydrodynamical disc simulation to examine its stability and creation. If the
vortex interpretation is correct, it could be the opening of a new era of
observational vortex research. The next generation of telescopes such as ALMA
(see Wolf \& Klahr \cite{wolf+klahr02}) will be able to probe protoplanetary
discs all over the sky for evidence of vortex activity. This will be the
ultimate test of whether the research done so far in the field has been just an
intellectual exercise, or if planets do indeed form partially as a result of
long-lived vortices. This would make the vortex dust-trapping theory an
important step in planet formation. Perhaps we owe the existence of our own
planet to vortices that were present in the solar nebular 4.6 billion years ago.

\begin{acknowledgement}
We would like to thank Pierre Barge, Hubert Klahr and Pierre-Henri 
Chavanis for inspiring discussions during the workshop ``Planetary formation:
toward a new scenario?'' held in Marseille in June 2003. We would also like to
thank the anonymous referee for constructive comments.
\end{acknowledgement}

\appendix

\section{Derivation of vortex solution}

In this Appendix we show how to construct an elliptical velocity field and an
appropriate corresponding enthalpy that satisfies \Eqs{eulgood}{contgood}. We
start out by proposing that the velocity field \Eqss{uxgood}{uzgood} is indeed
a solution, provided
that the aspect ratio and the angular velocity fulfil certain criteria. This
velocity field is divergence-free and has no $z$-component. It can then be
written as the curl of a stream function $\vec{\varPsi}$ with only a
$z$-component, $\vec{u} = \nab \times \vec{\varPsi}$.

Given the velocity field $\vec{u}$ one can attempt to construct an enthalpy such
that the flow is steady with $\partial \vec{u}/\partial t =
\vec{0}$. This implies that
\EQ
  \nab h = \varOmega^2 (3 \vec{x} - \vec{z}) - 2 \vec{\varOmega}
  \times \vec{u} - (\vec{u} \cdot \nab) \cdot \vec{u} \, .
\EN
The tidal force term and the vertical gravity term on the right hand side are
obviously gradient terms. The same is true for the Coriolis term since
\EQ
  \hat{\vec{k}} \times \vec{u} = \pmatrix{ 0 \cr 0 \cr 1
  } \times \pmatrix{ \partial \varPsi / \partial y
  \cr - \partial \varPsi / \partial x \cr 0 } = \pmatrix{
  \partial \varPsi / \partial x \cr \partial \varPsi / \partial y \cr 0 }
  = \nab \varPsi \, .
\EN
We are now left with (ignoring the integration constants for now)
\EQ
  \nab h = \nab ({\textstyle\frac{3}{2}}\varOmega^2 x^2 - {\textstyle\frac{1}{2}}\varOmega^2
  z^2 - 2 \varOmega \varPsi ) - (\vec{u} \cdot \nab) \cdot \vec{u} \, .
\EN
The advective term $(\vec{u} \cdot \nab) \vec{u}$ is a bit more tricky to
put in gradient form. It can be done by applying the vector identity
\EQ
  (\vec{u} \cdot \nab)\vec{u} = \nab ({\textstyle\frac{1}{2}}\vec{u}^2) -
  \vec{u} \times (\nab \times \vec{u}) \, .
  \label{advterm}
\EN
The first term on the right hand side comes out on gradient form. The
second can be calculated by noting that
\EQ
  \nab \times \vec{u} \equiv \vec{\omega} =
  \pmatrix{0 \cr 0 \cr - \partial^2\varPsi / \partial x^2 - \partial^2 
  \varPsi / \partial y^2} \, ,
\EN
where the vorticity $\vec{\omega}$ of the flow is introduced. The
second term on the right hand side of \Eq{advterm} can now be
rewritten as
\EQ
  \vec{u} \times (\nab \times \vec{u}) = 
  \pmatrix{ \partial \varPsi / \partial y \cr - \partial \varPsi / \partial x
  \cr 0 } \times \pmatrix{0 \cr 0 \cr \omega }
  = - \omega \nab \varPsi \, .
\EN
If the vorticity is constant and independent of the spatial
coordinates, this can be written
\EQ
  \vec{u} \times (\nab \times \vec{u}) =  \nab (- \omega \varPsi) \, .
  \label{constvort}
\EN
The gradient of the enthalpy can now be written entirely as a sum of
gradient terms,
\EQ
  \nab h = \nab ({\textstyle\frac{3}{2}}\varOmega^2 x^2 - {\textstyle\frac{1}{2}}\varOmega^2
  z^2 - 2 \varOmega \varPsi - {\textstyle\frac{1}{2}}\vec{u}^2 - \omega
  \varPsi) \, .
\EN
This is easily integrated to give
\EQ
  h = {\textstyle\frac{3}{2}}\varOmega^2 x^2 - {\textstyle\frac{1}{2}}\varOmega^2
  z^2 - (2 \varOmega + \omega) \varPsi - {\textstyle\frac{1}{2}}\vec{u}^2
  + {\rm const} \, ,
  \label{enth}
\EN
where all the integration constants are collected in just one constant.

In order to calculate the enthalpy of the flow given by \Eqss{uxgood}{uzgood} we need to
know also the stream function and the vorticity of the flow. The magnitude of
the stream function is
\EQ
  \varPsi = {\textstyle\frac{1}{2}}\left(\frac{1}{\epsilon} x^2 +
  \epsilon y^2\right) \varOmega' \, ,
\EN
and the vorticity has the magnitude
\EQ
  \omega = -\frac{\partial^2 \varPsi}{\partial x^2} -
  \frac{\partial^2 \varPsi}{\partial y^2} = -\left(\frac{1}{\epsilon} + \epsilon\right) \varOmega' \, .
\EN
This is a constant, and \Eq{enth} can therefore be applied.
To simplify the calculations we write the enthalpy as $h(x,y,z) = h_x(x)
+ h_y(y) + h_z(z)$ (as there are no mixed terms). For $h_x$ we get
\EQ
  h_x=({\textstyle\frac{3}{2}}\varOmega^2 - \varOmega \frac{1}{\epsilon} \varOmega'
          + {\textstyle\frac{1}{2}} \varOmega'^2) x^2 \equiv B x^2 \, ;
  \label{coeffB}
\EN
similarly for $h_y$,
\EQ
  h_y = (-\varOmega \epsilon \varOmega' + {\textstyle\frac{1}{2}}\varOmega'^2)
  y^2 \equiv A y^2 \, ;
  \label{coeffA}
\EN
and for $h_z$,
\EQ
  h_z = - {\textstyle\frac{1}{2}}\varOmega^2 z^2 \equiv C z^2 \, .
  \label{coeffC}
\EN
Now the enthalpy can be written as $h = B x^2 + A y^2 + C
z^2$. This can be done for any angular velocity $\varOmega'$ of the vortex.

An equilibrium flow solution must also have $\partial\rho/\partial t=0$
everywhere. The continuity equation \Eq{contgood} can be rewritten as
\EQA
  \frac{\partial \rho}{\partial t} + \nab \cdot (\rho \vec{u}) &=&
  \frac{\partial \rho}{\partial t} + \rho (\nab \cdot \vec{u}) + \vec{u} \cdot
  \nab \rho \nonumber \\ 
  &=& \frac{\partial \rho}{\partial t} + \vec{u} \cdot \nab
  \rho = 0 \, ,
\ENA
since the velocity field has $\nab \cdot \vec{u}=0$. This means that the
gradient of the density must everywhere be perpendicular to the flow, since
$\vec{u} \cdot \nab \rho$ must be equal to zero. In the absence of an entropy
gradient the gas is barotropic, so $\nab \rho$ and $\nab h$ are parallel,
and therefore $\vec{u} \cdot \nab h=0$.
This means that the contours of enthalpy must be ellipses with the same aspect
ratio as the vortex.

For the enthalpy given by $h(x,y,z)=B x^2+A
y^2 + C z^2$, where the coefficients are specified in \Eqss{coeffB}{coeffC},
and a velocity field given by \Eqss{uxgood}{uzgood}, this leads to
\EQ
  \frac{A}{B} = \epsilon^2 = 
  \frac{-\varOmega \epsilon \varOmega' + \frac{1}{2} \varOmega'^2}
  {\frac{3}{2} \varOmega^2 - \varOmega \frac{1}{\epsilon} \varOmega' + 
  \frac{1}{2} \varOmega'^2} \, ,
\EN
which implies that
\EQA
  {\textstyle\frac{3}{2}}\varOmega^2 \epsilon^2 - \varOmega \epsilon \varOmega' +
  {\textstyle\frac{1}{2}}\epsilon^2 \varOmega'^2 &=&
                -\varOmega \epsilon \varOmega' + {\textstyle\frac{1}{2}}\varOmega'^2 \nonumber \\
  \Rightarrow \varOmega' &=& \frac{\sqrt{3} \epsilon
                      \varOmega}{\sqrt{1-\epsilon^2}} \equiv \alpha
                      \varOmega \, .
\ENA
This solution requires that $0 \leqslant \epsilon < 1$. Negative $\epsilon$ in
the same range in principle also give solutions, but the enthalpy and velocity
field do not change since $\epsilon$ only enters as $\epsilon^2$ when
$\varOmega'$ is inserted. The
parameters $B$, $A$ and $C$ are
\EQA
  B &=& -{\textstyle\frac{1}{2}}\varOmega^2 \left(-\frac{3}{1-\epsilon^2} +   
  \frac{2\sqrt{3}}{\sqrt{1-\epsilon^2}}\right) \\
  A &=& -{\textstyle\frac{1}{2}}\varOmega^2
  \left(-\frac{3}{1-\epsilon^2}+\frac{2\sqrt{3}}{\sqrt{1-\epsilon^2}}\right)
  \epsilon^2 \\
  C &=& -{\textstyle\frac{1}{2}}\varOmega^2
\ENA
For $0.5 < \epsilon < 1$ the coefficients $B$ and $A$ are positive, resulting in a
region of low pressure with a clockwise rotation around it, contrary to the
counter-clockwise rotation around low-pressures on Earth. As in the GNG
paper we focus only on $0 \leqslant \epsilon \leqslant 0.5$.
Here $B$ and $A$ are negative, and the result is a high-pressure region.
Defining $\delta^2 = -\frac{3}{1-\epsilon^2} + \frac{2\sqrt{3}}{\sqrt{1-
\epsilon^2}}$ and requiring that the enthalpy vanish on the vortex boundary
gives
\EQ
  h = {\textstyle\frac{1}{2}}\delta^2 \varOmega^2 (b^2-x^2-\epsilon^2 y^2) - {\textstyle\frac{1}{2}}
  \varOmega^2 z^2 \, .
\EN

\end{document}